\documentstyle[aps,12pt]{revtex}

\begin{document}
\draft
\title{Nearest-Neighbor Correlations in Hubbard Model}
\author{Yu.B.Kudasov$\thanks{%
E-mail: kudasov@ntc1.vniief.ru}$}
\address{Russian Federal Nuclear Center - VNIIEF\\
pr. Myra 37, Sarov , Nizhni Novgorod\\
Region, Russia, 607190}
\date{\today}
\maketitle

\begin{abstract}
The Hubbard Hamiltonian is investigated by means of a new variational trial
wave function. The trial wave function includes either intrasite and nearest
- neighbor correlations in an explicit form. To calculate density matrices
the method of Kikuchi's pseudoensemble is used. The case of half-filled
fermionic band is carefully investigated in the limit of a large number of
lattice sites. The ground state energy and correlation functions are
determined for Bethe lattices with $z=2,4$ and $6$ nearest neighbors.
\end{abstract}

\pacs{PACS: 71.10.Fd, 71.10.Hf, 71.27.+a, 71.28+d}

\begin{abstract}
Key words: strongly correlated fermions, trial wave function, nearest
neighbor correlations
\end{abstract}

\newpage

Short-range correlations have been shown to have a considerable influence on
properties of strongly interacting fermions. In the most refined way, the
problem is stated within the framework of the one-band Hubbard model \cite
{Hubbard,Gutzw,Kanam}. If fermions of spin $1/2$ hop to the adjacent sites
of a lattice only, the Hubbard Hamiltonian has the following form 
\begin{equation}
H=H_{k}+H_{p}=t\sum_{\left\langle ij\right\rangle ,\sigma }\left( a_{i\sigma
}^{\dagger }a_{j\sigma }+H.c.\right) +U\sum_{i}n_{i\uparrow }n_{i\downarrow }
\label{Hubb}
\end{equation}
where $a_{i\sigma }^{\dagger }\left( a_{j\sigma }\right) $ is the creation
(annihilation) operator of a fermion of spin $\sigma $ on the $i$-th lattice
site, $\left\langle ij\right\rangle $ denotes a pair of adjacent sites, $%
n_{i\sigma }=a_{i\sigma }^{\dagger }a_{j\sigma }$.

There are known few exact solutions of the Hamiltonian (\ref{Hubb}): for a
homogeneous chain (1D) \cite{Lieb}, so-called Gutzwiller approach (GA) for a
lattice of infinitive dimensions D$=\infty $ \cite{Gutzw,Metz} and some
other special cases \cite{Kolo,Brandt}. Lattices of intermediate dimensions,
especially 2D and 3D ones, are of great practical importance, and a number
of works is devoted to their investigation (see \cite{Georges,Senatore} and
reference therein). In particular, the Gutzwiller's trial wave function
which is exact in the limit of D$=\infty $ was applied \cite
{Yokoy,Metz2,Gebh}. This wave function entered in a numerical procedure of
the variational Monte Carlo (VMC) method \cite{Yokoy} or was used in
diagrammatic expansions of GA in terms of $1/$D \cite{Metz2,Gebh}. In the
last case, one had to base on a strong assumption that the ground state of
2D and 3D lattices only slightly differed from that of the D$=\infty $
lattice. Actually, the problem of lattices of finite dimensions lies in
necessity of inclusion of spacial correlations, especially short-range ones.
As was shown phenomenologically in the framework of Nearly Antiferromagnetic
Fermi Liquid theory \cite{Pines}, in some situations the short-range
correlations are very strong and, therefore, a microscopic non-perturbative
approach is needed to treat them.

In the present Letter, we explore the Hamiltonian (\ref{Hubb}) with
half-filled initial fermionic band by means of a trial wave function of
Gutzwiller's type which, in addition to intrasite correlations, includes
nearest-neighbor ones in an explicit form. Let us consider a system of $N$
fermions of spin $1/2$ on a lattice consisted of $L$ sites. Then, in a
general form, the correlated $N$-particle trial wave function can be written
as 
\begin{equation}
\left| \psi \right\rangle =\prod_{\lambda }g_{\lambda }^{\widehat{P}%
_{\lambda }}\left| \varphi _{0}\right\rangle  \label{GTW}
\end{equation}
where $\left| \varphi _{0}\right\rangle $ is the $N$-particle wave function
of free fermions, for instance the Fermi sea $\prod_{k<k_{F\uparrow
}}a_{k\uparrow }^{\dagger }$ $\prod_{k<k_{F\downarrow }}a_{k\downarrow
}^{\dagger }\left| 0\right\rangle $, $k$ is the wave number of a fermion, $%
k_{F\sigma }$ is the Fermi wave number of fermions of spin $\sigma $, $%
g_{\lambda }$ are the parameters which lies within interval $\left] 0,\infty
\right[ $, $\widehat{P}_{\lambda }$ are the projection operators onto all
feasible configurations of a lattice site and a pair of adjacent sites.
There are 4 such operators for itrasite configurations 
\begin{equation}
\widehat{X}_{1}=\sum_{i}\left( 1-n_{i\uparrow }\right) \left(
1-n_{i\downarrow }\right) ,\widehat{X}_{2}=\sum_{i}n_{i\uparrow }\left(
1-n_{i\downarrow }\right) ,\widehat{X}_{3}=\sum_{i}\left( 1-n_{i\uparrow
}\right) n_{i\downarrow },\widehat{X}_{4}=\sum_{i}n_{i\uparrow
}n_{i\downarrow }\text{,}  \label{x}
\end{equation}
and 10 operators for nearest neighbor configurations, 
\begin{equation}
\widehat{Y}_{1}=\sum_{\left\langle ij\right\rangle }\left( 1-n_{i\uparrow
}\right) \left( 1-n_{i\downarrow }\right) \left( 1-n_{j\uparrow }\right)
\left( 1-n_{j\downarrow }\right) ,\widehat{Y}_{2}=\sum_{\left\langle
ij\right\rangle }n_{i\uparrow }n_{i\downarrow }n_{j\uparrow }n_{j\downarrow }%
\text{, and etc.}  \label{y}
\end{equation}
(see Table \ref{table}).

The trial wave function (\ref{GTW}) remains to be antisymmetric and
conserves translational properties of $\left| \varphi _{0}\right\rangle $.
On the other hand, either intrasite and nearest-neighbor correlations are
taken into consideration explicitly. From now on, we shall consider lattices
for which the total number of nearest-neighbors pairs is equal to $zL/2$,
where $z$ is the number of nearest neighbors of a site. Let us denote
normalized eigenvalues of the projection operators as $x_{\lambda }\left|
\Phi \right\rangle =L^{-1}\widehat{X}_{\lambda }\left| \Phi \right\rangle $, 
$y_{\lambda }\left| \Phi \right\rangle =\left( zL/2\right) ^{-1}\widehat{Y}%
_{\lambda }\left| \Phi \right\rangle $. The eigenvalues turn out to be
related to each other by normalization conditions \cite{Ziman} 
\begin{equation}
\sum_{\lambda }x_{\lambda }=1\text{, }\sum_{\lambda }\beta _{\lambda
}y_{\lambda }=1  \label{norm}
\end{equation}
and self-consistency conditions \cite{Ziman} 
\begin{eqnarray}
y_{1}+y_{3}+y_{4}+y_{5} &=&x_{1}\text{,}  \label{self} \\
y_{2}+y_{3}+y_{8}+y_{9} &=&x_{4}\text{,}  \nonumber \\
y_{4}+y_{6}+y_{7}+y_{8} &=&x_{2}\text{,}  \nonumber \\
y_{5}+y_{7}+y_{9}+y_{10} &=&x_{3}\text{.}  \nonumber
\end{eqnarray}

As concentrations of fermions of each spins are fixed there are the only
independent parameter $x_{\lambda }$ and 7 independent parameters $%
y_{\lambda }$. In the case of half-filled band, additional constrains appear 
\begin{equation}
y_{1}=y_{2}\text{, }y_{6}=y_{10}\text{, }y_{4}=y_{5}=y_{8}=y_{9}  \label{add}
\end{equation}
which reduce the number of independent parameters $y_{\lambda }$ to 3. It
should be noted that, in contrast to (\ref{norm}) and (\ref{self}), the
condition (\ref{add}) imposes the restriction on averaged values of $y_{i}$
only, that is, configurations, which violate (\ref{add}), enter into the
trial wave function (\ref{GTW}). Nevertheless, as will be shown below, one
can omit them in the limit $L\longrightarrow \infty $.

Assume that $x_{1}=x_{4}=x$, $y_{3}$, $y_{4}$, and $y_{7}$ are independent
parameters. Then, taking into account additional degeneracy of the
half-filled band, the correlated trial wave function is transformed to

\begin{equation}
\left| \psi \right\rangle =g_{0}^{\widehat{X}}g_{3}^{\beta _{3}\widehat{Y}%
_{3}}g_{4}^{4\beta _{4}\widehat{Y}_{4}}g_{7}^{\beta _{7}\widehat{Y}%
_{7}}\left| \varphi _{0}\right\rangle  \label{CWT}
\end{equation}

At first, we calculate the norm of the correlated trial wave function 
\begin{equation}
\left\langle \psi \mid \psi \right\rangle =\sum_{\left\{
x,y_{3},y_{4},y_{7}\right\} }W_{\left\{ x,y_{3},y_{4},y_{7}\right\}
}g_{0}^{2Lx}g_{3}^{2zLy_{3}}g_{4}^{8zLy_{4}}g_{7}^{2zLy_{7}}=\sum_{\left\{
x,y_{3},y_{4},y_{7}\right\} }R_{\left\{ x,y_{3},y_{4},y_{7}\right\} }
\label{norma}
\end{equation}
where the set $\left\{ x,y_{3},y_{4},y_{7}\right\} $ describes
configurations which contains $Lx$ doubly occupied sites of the lattice, $%
zLy_{3}/2$ nearest neighbor pairs of $\widehat{Y}_{3}$ type and etc. A
unessential factor is omitted in (\ref{norma}) for simplicity. The summation
is extended over all possible sets $\left\{ x,y_{3},y_{4},y_{7}\right\} $. $%
W_{\left\{ x,y_{3},y_{4},y_{7}\right\} }$ is the number of configurations
with the fixed set of the independent parameters. To calculate $W_{\left\{
x,y_{3},y_{4},y_{7}\right\} }$ we shall use the Kikuchi's pseudo-ensemble
method \cite{Ziman,Kik}. It should be mentioned that this method is exact on
the Bethe lattice and approximated for lattices with closed paths \cite
{Ziman}. According to Kikuchi's hypothesis, we have

\begin{equation}
W=\Gamma Q\text{, }Q=\frac{\left( zL/2\right) !}{\prod_{\lambda }\left[
\left( zy_{\lambda }L/2\right) !\right] ^{\beta _{i}}}\text{, }\Gamma =\frac{%
L!\prod_{\lambda }\left( x_{\lambda }zL\right) !}{\left( zL\right)
!\prod_{\lambda }(x_{\lambda }L)!}  \label{kikuchi}
\end{equation}
where lower indexes of $W$, $\Gamma $, and $Q$ are omitted for the sake of
simplicity. $Q$ determines a number of arrangements of 10 elements
corresponding to $\widehat{Y}_{\lambda }$ taken $zL/2$ at a time. $\Gamma $
is a fraction of correct arrangements in the pseudo-ensemble. In Eq.(\ref
{kikuchi}), dependent $x_{\lambda }$ and $y_{\lambda }$ should be expressed
in terms of $x$, $y_{3}$, $y_{4}$, and $y_{7}$. In the usual fashion, in the
thermodynamic limit $L\rightarrow \infty $, we retain only the terms of the
series (\ref{norma}) which are very close to the largest one. As far as $%
R_{\left\{ x,y_{3},y_{4},y_{7}\right\} }$ is a positive function, one can
search a maximum of its logarithm instead of the function itself. Let us
retain the main terms on $L$ only after taking the logarithm. It can be
shown that this approach corresponds to substitution $(zL/2)!\longrightarrow
(L!)^{z/2}$ which was usually used \cite{Ziman,Kik}. Then we obtain 
\begin{equation}
\frac{\ln W}{L}=2\left( z-1\right) \left[ x\ln x+\left( \frac{1}{2}-x\right)
\ln \left( \frac{1}{2}-x\right) \right] -z\left[ y_{2}\ln y_{2}+y_{3}\ln
y_{3}+4y_{4}\ln y_{4}+y_{6}\ln y_{6}+y_{7}\ln y_{7}\right]  \label{lnW}
\end{equation}
where $y_{2}=x-y_{3}-2y_{4}$, $y_{6}=1/2-x-y_{7}-2y_{4}$. The domain of
function (\ref{lnW}) where we search the maximum is limited by conditions (%
\ref{norm}) and (\ref{self}). At its boundaries the gradient of function (%
\ref{lnW}) is directed inwards the domain. That is why, the global maximum
of $\ln R/L$ should be an inner one, and conditions $\partial \ln R/\partial
\eta _{\lambda }=0$ where $\eta _{\lambda }=x,y_{3},y_{4},y_{7}$ are
necessary for the global maximum. They lead to the following system of
equations which relate $g_{i}$ to $x$ and $y_{i}$

\begin{eqnarray}
g_{0} &=&\left( \frac{1/2-x}{x}\right) ^{z-1}\left( \frac{x-y_{3}-2y_{4}}{%
1/2-x-y_{7}-2y_{4}}\right) ^{z/2},  \label{sys} \\
g_{3}^{2} &=&\frac{y_{3}}{x-y_{3}-2y_{4}},  \nonumber \\
g_{4}^{4} &=&\frac{y_{4}^{2}}{\left( 1/2-x-y_{7}-2y_{4}\right) \left(
x-y_{3}-2y_{4}\right) },  \nonumber \\
g_{7}^{2} &=&\frac{y_{7}}{1/2-x-y_{7}-2y_{4}}.  \nonumber
\end{eqnarray}

To determine the expectation value of the Hamiltonian (\ref{Hubb}) we need
to calculate a density matrix of the first order:

\begin{equation}
\rho _{1}=\frac{1}{L}\frac{\left\langle \psi \right|
\sum\limits_{<ij>,\sigma }\left( a_{i\sigma }^{\dagger }a_{j\sigma
}+H.c.\right) \left| \psi \right\rangle }{\langle \psi \left| \psi
\right\rangle }.  \label{density}
\end{equation}

In contrast to GA, while a fermion hops from site $i$ to $j,$ it should be
taken into account that configurations of adjacent pairs $k-i$ and $i-l$
change (Fig.\ref{fig3}). Let us fix a particular lattice fragment (Fig.\ref
{fig3}), and calculate function $W^{\prime }$ of the remain lattice from
Eqs.(\ref{kikuchi}). Then, a fraction of configurations, which contain the
fragment, can be found as

\begin{equation}
\frac{W^{\prime }}{W}=\frac{y_{(ij)}\prod\limits_{k}y_{(ki)}\prod%
\limits_{l}y_{(jl)}}{x_{(i)}x_{(j)}}.  \label{Wprim}
\end{equation}

Using (\ref{Wprim}), we sum up all contributions to the density matrix.
Then, it takes the following form

\begin{equation}
\rho _{1}=4\left[ 2y_{4}\left( a_{1}a_{2}\right) ^{z-1}+\frac{y_{3}g_{7}}{%
g_{0}g_{3}}a_{1}^{2(z-1)}+\frac{y_{7}g_{0}g_{3}}{g_{7}}a_{2}^{2(z-1)}\right]
,  \label{ro}
\end{equation}
where $a_{1}=\left( y_{2}g_{4}+y_{3}g_{4}/g_{3}+y_{4}\left( g_{7}+1\right)
/g_{4}\right) /x$ and $a_{2}=\left( y_{6}g_{4}+y_{7}g_{4}/g_{7}+y_{4}\left(
g_{3}+1\right) /g_{4}\right) /(1/2-x)$. The first term of (\ref{ro})
describes a motion of a fermion in the Hubbard subbands. The second and
third ones arise from transitions between the subbands. In this expression,
parameters $g_{0},g_{3},g_{4},g_{7}$ should be excluded by means of system (%
\ref{sys}). After some simplifications we find

\begin{equation}
\rho _{1}=8\left( y_{4}+\sqrt{y_{3}y_{7}}\right) \left[ \frac{y_{4}}{x(1/2-x)%
}\left( \sqrt{y_{2}}+\sqrt{y_{3}}+\sqrt{y_{6}}+\sqrt{y_{7}}\right)
^{2}\right] ^{z-1}.  \label{rof}
\end{equation}

Finally, the total energy can be presented in the Gutzwiller's form

\begin{equation}
E=\frac{1}{L}\frac{\left\langle \psi \left| H\right| \psi \right\rangle }{%
\left\langle \psi \mid \psi \right\rangle }=q\varepsilon _{0}+xU  \label{Ham}
\end{equation}
where $q=\rho _{1}/\rho _{1}^{0}$, $\rho _{1}^{0}$ is the density matrix (%
\ref{rof}) at $U=0$, $\varepsilon _{0}$ is the average energy of the free
fermions. The ground state energy is determined as $\min_{\left\{
x,y_{3},y_{4},y_{7}\right\} }\left( E\right) $. The function $E$ turns out
to be smooth, and its minimum is easily found numerically. It also should be
noted that the total energy is invariant in respect of the following
substitution: $U\rightarrow -U,y_{7}\longleftrightarrow
y_{3},g_{7}\longleftrightarrow g_{3},g_{0}\longleftrightarrow 1/g_{0}$. A
detailed discussion of the method presented above will be published
elsewhere.

As was mentioned above, the method used is exact for a Bethe lattice. We
consider our results for $z=4$ and $6$ as approximate solutions for 2D and
3D lattices correspondingly. They are shown in the Fig.\ref{fig1}, together
with that of 1D chain, and compared with the exact solutions for 1D chain 
\cite{Lieb}, D$=\infty $ lattice \cite{Gutzw}, and numerical results of VMC
method for a paramagnetic phase \cite{Yokoy}. The perturbative expansion (GA$%
+1/$D$+1/$D$^{2}$) is not shown because it is very close to the GA curve for
2D and 3D (a shift of critical values of $U$ between GA and GA$+1/$D$+1/$D$%
^{2}$ is a few percent only \cite{Gebh}). To investigate the D$=\infty $
limit of our model we carried out calculations for $z=50,100,200$. They show
that the ground state energy tends to the exact D$=\infty $ solution while $%
z $ (or D) increasing. As one can see from the Fig.\ref{fig1}, at
intermediate interaction of fermions ($U\sim 1$), our method gives the
ground state energy significantly lower than that of VMC or GA$+$1/D$+$1/D$%
^{2}$ procedures.. This means that nearest-neighbor correlation are
essential in this region.

Symmetric and antisymmetric correlation functions of the nearest neighbor 
\begin{eqnarray}
G_{s} &=&\left\langle n_{\uparrow }n_{\uparrow }\right\rangle ^{\prime
}+\left\langle n_{\downarrow }n_{\downarrow }\right\rangle ^{\prime
}=2\left( y_{2}+2y_{4}+y_{6}\right) ,  \label{corr} \\
G_{a} &=&\left\langle n_{\uparrow }n_{\downarrow }\right\rangle ^{\prime
}+\left\langle n_{\downarrow }n_{\uparrow }\right\rangle ^{\prime }=2\left(
y_{2}+2y_{4}+y_{7}\right)  \nonumber
\end{eqnarray}
are shown in Fig.\ref{fig2} for the same lattices as in Fig.\ref{fig1}. The
prime in Eqs.\ref{corr} denotes the averaging over nearest-neighbor pairs
only. One can see in Fig.\ref{fig2} that even at $U=0$ some correlations
appear due to a exchange hole. An increasing of $U$ leads to enhancing of
correlations but there exists a saturation point. At further increasing of $%
U $ the nearest-neighbor correlations remain almost constant and variation
of the ground state energy is due to intrasite correlations only.

In conclusion, a new non-perturbative approach to problem of strongly
correlated fermions is reported. A trial wave function which includes
nearest-neighbor correlations is constructed. For a half-filled initial
fermionic band, the ground state energy of the Hubbard Hamiltonian as well
as correlation functions are calculated for 2D and 3D lattices and the 1D
chain. Since the correlated wave function $\left| \psi \right\rangle $ has
the same translational properties as $\left| \varphi _{0}\right\rangle $
does, all the results obtained above describe a paramagnetic state. To build
entire phase diagram for Hubbard model our approach need to be extended to
an antiferromagnetic phase. The method may be especially beneficial when
systems with strong short-range correlations, for example CuO$_{2}$ planes
of HSTC, is considered.

The author thanks Prof. J.Brooks and Dr. W.Lewis for their continuous
encouragement. The work was supported by the International Science and
Technology Center under Project \#829.

\newpage 
\begin{figure}[tbp]
\caption{A fragment of a tree $(z=4)$. While a fermion hops from $j$ to $k$
site, configurations of adjacent pairs need to be taken into account.}
\label{fig3}
\end{figure}

\begin{figure}[tbp]
\caption{The ground state energy obtain by minimization of Eq.(17) for 1D
chain ($z=2$), $z=4$ (2D), and $z=6$ (3D) lattices (solid lines); the exact
solutions: GA ( dotted line) [2], 1D chain (dash-dot line) [4]; and
numerical VMC calculations (dash line).}
\label{fig1}
\end{figure}

\begin{figure}[tbp]
\caption{The correlation functions $G_{s}$ (dashed lines) and $G_{a}$ (solid
lines) of 1D chain, $z=4$ (2D) and $z=6$ (3D) lattices.}
\label{fig2}
\end{figure}

\begin{table}[tbp]
\caption{Pair projection operators, corresponding configurations and the
degeneracy factor}
\label{table}%
\begin{tabular}{dddddd}
Operator  &\multicolumn{2}{c}{Configuration}& Degeneracy \\
$\widehat{Y}_{i}$ &Site A&Site B& $\beta_{i}$\\
\tableline
$\widehat{Y}_{1}$ &   & &1 \\
$\widehat{Y}_{2}$& $\uparrow$ $\downarrow$& $\uparrow$ $\downarrow$ &1\\
$\widehat{Y}_{3}$& $\uparrow$ $\downarrow$&  &2\\
$\widehat{Y}_{4}$& $\uparrow$ & &2\\
$\widehat{Y}_{5}$& $\downarrow$ & &2\\
$\widehat{Y}_{6}$& $\uparrow$ &$\uparrow$ &1\\
$\widehat{Y}_{7}$& $\uparrow$ &$\downarrow$ &2\\
$\widehat{Y}_{8}$& $\uparrow$ $\downarrow$ &$\uparrow$ &2\\
$\widehat{Y}_{9}$& $\uparrow$ $\downarrow$ &$\downarrow$&2\\
$\widehat{Y}_{10}$& $\downarrow$& $\downarrow$ &1\\
\end{tabular}
\end{table}

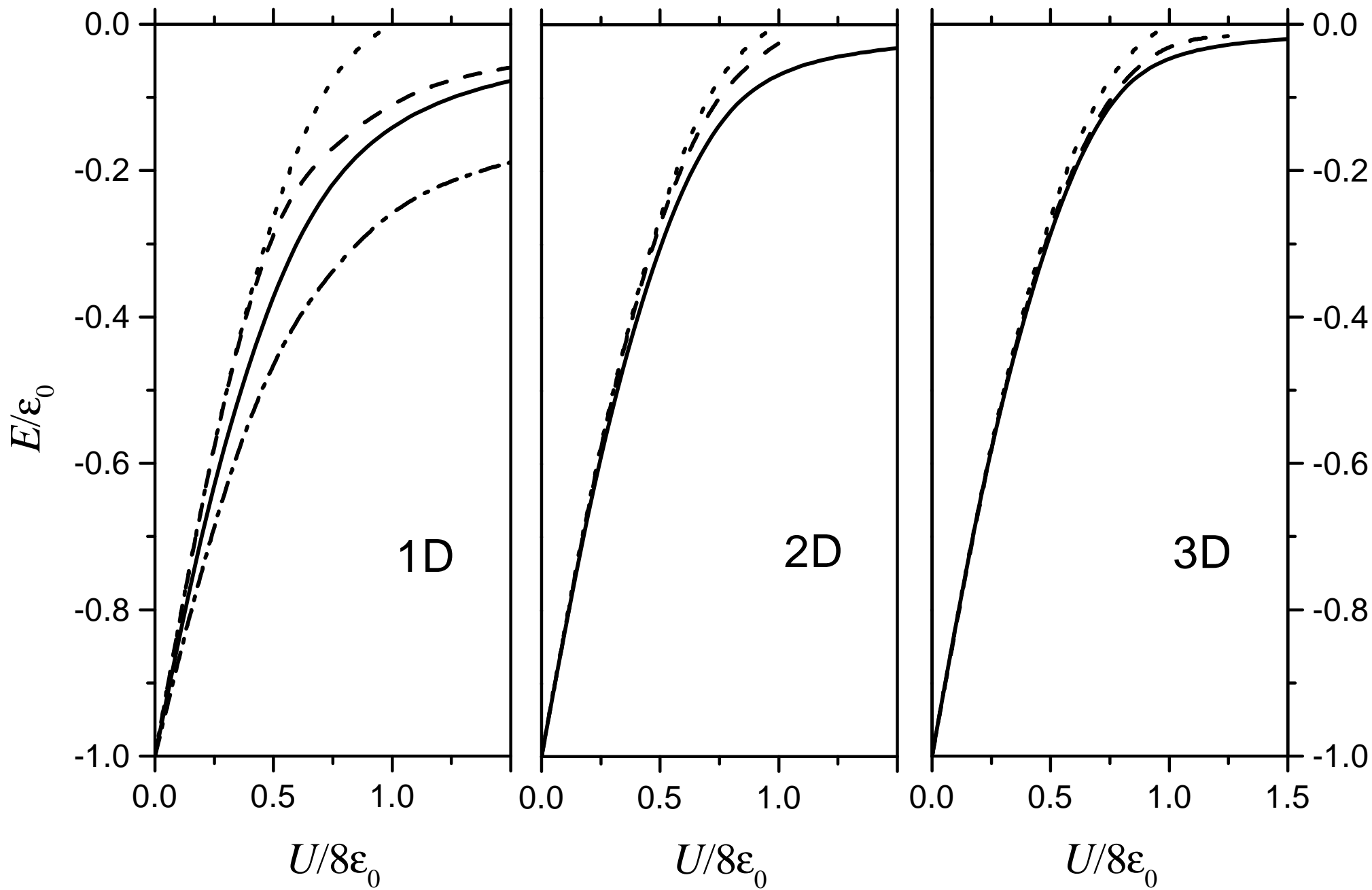
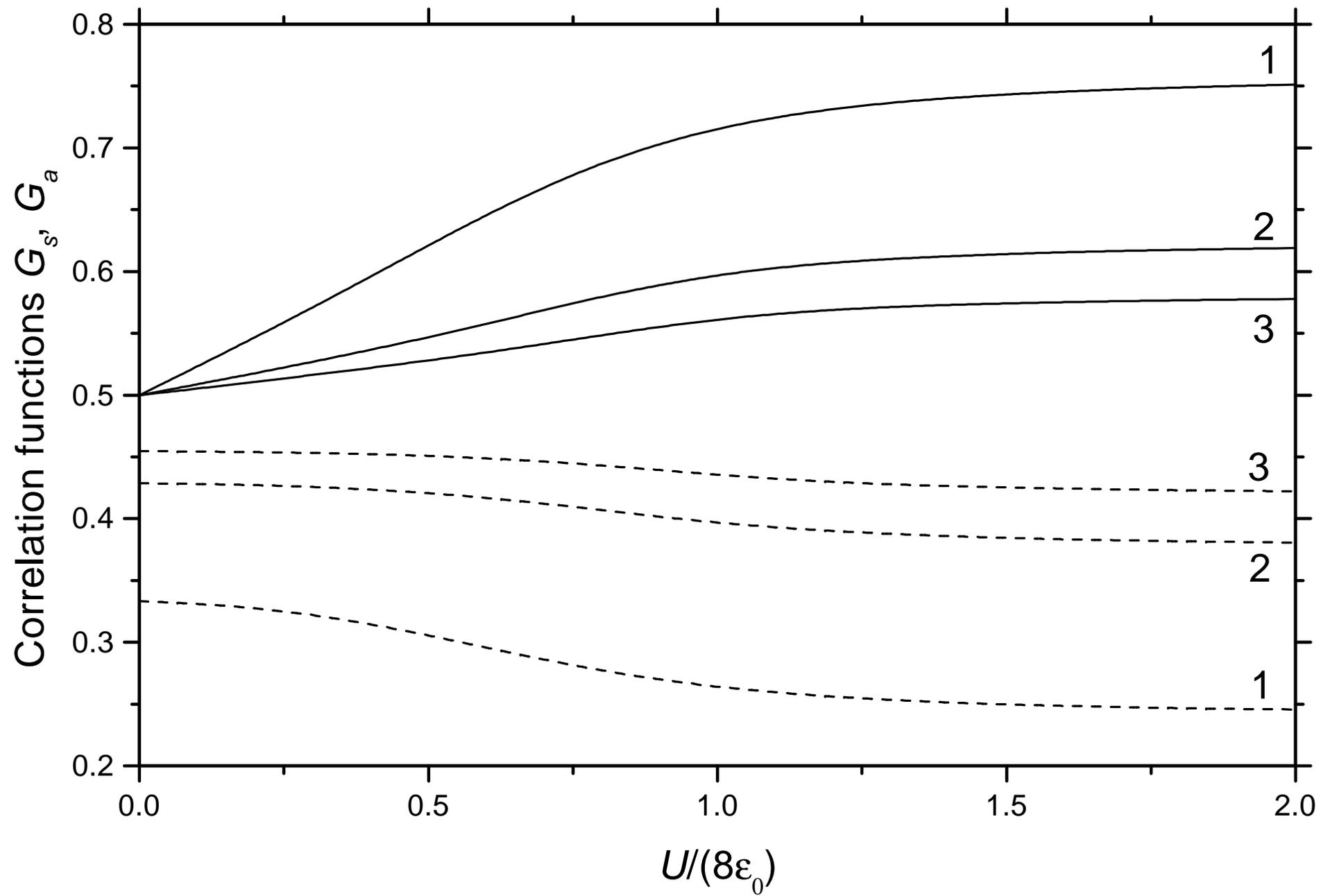
\end{document}